\documentstyle{article}

\newcommand{\mathbf}{\bf}

\begin{document}

\begin{center}
{\huge\bf A Topological Approach to Quantum Electrodynamics}
\end{center}

\vspace{1cm}
\begin{center}
{\large\bf
F.GHABOUSSI}\\
\end{center}

\begin{center}
\begin{minipage}{8cm}
Department of Physics, University of Konstanz\\
P.O. Box 5560, D 78434 Konstanz, Germany\\
E-mail: ghabousi@kaluza.physik.uni-konstanz.de
\end{minipage}
\end{center}

\vspace{1cm}

\begin{center}
{\large{\bf Abstract}}
\end{center}

\begin{center}
\begin{minipage}{12cm}
The physical reasons in favour of a two dimensional topological  
model of quantum electrodynamics are discussed. It is shown that in  
accord with this model there is a new uncertainty relation for  
photon which is compatible with QED.
\end{minipage}
\end{center}

\newpage
We discuss a two dimensional topological approach to quantum  
electrodynamics which can be helpful to understand dynamical aspects  
of two dimensional topological quantum effects such as flux  
quantization, cyclotron motion, Aharonov-Bohm effect and QHE. Note  
that all these effects can be considered as two dimensional quantum  
electrodynamical phase effects caused by a magnetic field which is  
prependicular to the two dimensional surface of motion of electrons.

Our motivation in considering such a theory is based on the  
physical fact that the quantum field of electromagnetic interaction,  
the photon, possess only two degrees of freedom which refers to a  
two dimensional geometrical background. Thus even in the four  
dimensional Maxwell theory one reduces the original four degrees of  
freedom of the electromagnetic field to the two physical degrees of  
freedom, e. g. by the use of radiation gauge.
Further it is based on the phenomenological fact that the mentioned  
two dimensional quantum phase, i. e. $e \oint\limits_{(contour)}  
A_m dx^m = N h; m,n = 1, 2; N \in {\mathbf Z}$ where $e$ and $A_m$  
are the electric charge and the electromagnetic potential, is  
varified by flux quantization and Aharonov-Bohm effect. It is based  
also on theoretical consideration that even the {\it four  
dimensional} Maxwell action: $\int F \wedge * F \;, \; F = F_{mn}  
dx^m \wedge dx^n$ must be described by the substructure of  
electromagnetic {\it two-forms} $F$ and $*F$.

Note also that in view of global character of quantum theory which  
is manifested through the globality of quantum state $( \sim$ wave  
function ), the global aspects of the underlying manifolds or bundle  
manifolds are essential for the structure of quantum theories which  
are defined on this manifold. In this sense topological invariants,  
e. g. cohomology, homology or harmonics on the mentioned manifolds  
determine global aspects of the mentioned quantum theories  
\cite{nak}. Accordingly we show that in view of the main role played  
by the second cohomology $H^2 ( M_{4D}) \cong Harm^2 ( M_{4D})$ of  
space-time four manifold in electrodynamics, such two-forms and  
their "dual" surfaces are essential even in four dimensional quantum  
electrodynamics. Hence with respect to topological quantum effects,  
two dimensional quantum theories defined by such two forms on such  
surfaces can replace four dimensional quantum theories.

We will show first that the structure of action and so the  
equations of motion of four dimensional Maxwell theory can be  
considered as conditions which restrict the electromagnetic  
two-forms to be defined on a two dimensional submanifold of the  
original four dimensional space-time. Recall that Maxwell equations  
in QED are considered also as conditions on the quantum state  
\cite{dirac}.

\bigskip
There are various local or differential and global or topological  
hints about the main role played by the two dimensional  
substructures of the four dimensional classical structure in four  
dimensional classical and quantum electrodynamics, i. e. about the  
restriction of the relevant structures in both theories to two-forms  
and two dimensional submanifolds. The first one is the absence of  
three- and {\it simple} four-forms in the Lagrangian of  
electrodynamics in view of its restriction to two-forms, altuough  
the underlying manifould is a four dimensional one. This fact alone  
can be considered as the irrelevance of higher than simple two-forms  
in electromagnetism, since if there were simple electromagnetic  
four or three-forms, they should be involved in such a general four  
dimensional theory. On the other hand to define a two form a two  
dimensional manifold is sufficient, so that an electromagnetic field  
strength can be defined on a two dimensional submanifold of the  
$(3+1)$- dimensional space-time. A reasonable two dimensional  
boundaryless non-boundary submanifold: $M_{2D} \in H_2 (M_{4D})$  
given as a part of three dimensional space seems to be suitable for  
our objection.

The second hint is that, accordingly, the vacuum equations of  
motions of both electrodynamics, i. e. $d F = 0$ and $d^{\dagger} F  
= 0$ restrict the involved two-forms which are assummed to be  
defined on a four manifold $M_{4D}$ to be harmonic forms: $F \in  
Harm^2 ( M_{4D})$. Thus the relevant electrodynamical two-form $F  
\in Harm^2 ( M_{4D})$ have no contribution from higher than two  
dimensional structures of the underlying four manifold, although the  
general possible electrodynamical two form on a compact orientable  
four manifold without boundary is given by the Hodge decomposition  
$F = d A \oplus d^{\dagger} \Omega^3 (A, F) \oplus Harm^2$  
\cite{bound}. Since, if one considers the term $d^{\dagger} \Omega^3  
(A, F)$ as the contribution of the higher than two dimensional  
structure of the four manifold to the structure of two forms, then  
the absence of this term shows that the relevant electromagnetic  
two-forms are those which {\it can} be defined only on a two  
dimensional submanifold $M_{2D}$ of the four manifold. Hence $F = d  
A \oplus Harm^2$ is an element of $H^2 ( M_{4D})$ which is  
isomorphic to $Harm^2 ( M_{4D})$. Note also that the dimension of  
the $H^2 ( M_{2D}) \cong H^2 ( M_{2D})$ is closely related with the  
invariant aspects of two dimensional manifolds.

The third hint is that in QED only slowly varying field strengths:  
$d F << F$ and $\partial_t F << F$ produce finite terms which  
enables one to renormalize QED \cite{weis}, i. e. again only fields  
with $d F = 0$ and in view of equations of motion also with  
$d^{\dagger} F = 0$ are QED relevant. It means that only the  
solutions of Maxwell equations $d F = 0, \ d^{\dagger} F = 0$  
produce physical, i. e. finite, results in QED. This is in accord  
with the path integral quantization idea where only the real  
classical path, which is the solution of classical equations of  
motion, contributes to the phase of quantum state.

As the last local hint let us mention that the electromagnetic  
field strength in the Landau gauge, which is the usual one for  
quantization in presence of magnetic field, is restricted to the two  
dimensional field which is defined on the {\it two dimensional}  
spatial submanifold: $F ( M_{2D})$ \cite{land}. Thus even the  
phenomenological description of magnetic quantization \cite{land} is  
obliged to use the two dimensional two-forms $F ( M_{2D})$ instead  
of the four dimensional ones: $F ( M_{4D})$.

In other words the four dimensional classical as well as quantum  
electrodynamics are not only based on the two dimensional  
substructure of two-forms, but both theories restrict these  
two-forms to be harmonic two-forms which should be defined on the  
$M_{2D}$ subsmanifolds of the four manifold.
Moreover recall that in QED only average of field strengths over  
finite space-time regions have a well defined meaning \cite{bohr}.  
Hence these averages can be considered as to be averaged over the  
rest $(1+1)$ dimensional part and to be defined only on $M_{2D}$  
submanifold of the $( 3 + 1 )$ dimensional space-time. Thus one can  
consider the averaged field strength to be constant with respect to  
the rest and to depend only on variables on the $M_{2D}$  
submanifold.

Also the topological invariant peoperties of Maxwell action $\int F  
\wedge * F$, which are essential with respect to the topological (  
global) character of quantum phases, refer to two dimensional  
invariants, since all relevant invariants here are constructed from  
the electromagnetic two form $F$:

The electromagnetic elements of the four manifold cohomology  
invariants are the action function $f_{(em)} (A(x)) = \int F (A)  
\wedge * F (A) ; \ f_{(em)} (A(x)) \in H^0 (M_{4D})$, the closed  
non-exact electromagnetic two-form $F \in H^2 (M_{4D})$ and the  
Maxwell-form $\Omega^4 (M_{4D}):= F \wedge * F; \ \Omega_{(em)} ^4  
(M_{4D}) \in H^4 (M_{4D})$ are constructed from the electromagnetic  
two-form $F = d A \oplus Harm^2 _{(em)}; \ F \in H^2 _{(em)}  
(M_{4D})$.
In other words $H^2 _{(em)}(M_{4D}) \times H^2 _{(em)} (M_{4D})  
\rightarrow H^4 _{(em)}  \cong H^0 _{(em)} (M_{4D})$. Note that in  
view of Hodge theorem on compact orientable Riemannian manifolds:  
$H^r (M) \cong Harm^r (M)$ and $dim H^r (M) = dim Harm^r (M)$ all  
above isomorphisms are given also between harmonics, so that the  
solutions of Maxwell equations $\in Harm^2 (M_{4D})$ determine in  
this way global aspects of $M_{4D}$ like the Euler characteristic  
$\chi (M_{4D}) = \Sigma (-1)^r dim Harm^r (M_{4D})$. Thus in the  
simply connected case of interest where $H^1 \cong H^3 = 0$ only  
$dim Harm^2$ will determine the Euler characteristic of our  
$M_{4D}$.
Therefore even the topological aspects of four dimensional QED:  
$H^4 \cong Harm^4 \cong Harm^0 \cong H^0$ which are essential in  
topological quantum effects are given by the two dimensional  
topological aspects, since these topological invariants of four  
manifold are given in terms of topological invariants of the two  
dimensional submanifold. Thus, if with regard to the Maxwell-form  
$H^2 _{(em)} (M_{4D}) \times H^2 _{(em)} (M_{4D}) \rightarrow H^4  
_{(em)} (M_{4D})$, also the "dual" map $H_2 (M_{4D}) \times H_2  
(M_{4D}) \rightarrow H_0 (M_{4D}) \cong H_4 (M_{4D})$ is given, then  
the homological invariants of our four manifold are also given by  
the invariants of its two dimensional submanifold: $M_{2D} \in H_2  
(M_{4D})$.

Moreover recall that the invariant of $F$, i. e.  
$\int\limits_{surface} F$, is obtained with respect to a two  
dimensional {\it surface} and also that in view of the absence of  
three-forms on two dimensional manifolds $M_{2D}$ any two-form on  
these manifolds is a closed two form, i. e. in our case $d F(M_{2D})  
= 0$.
Note also that the two dimensional equations of motion $d  
F^{\dagger} (M_{2D})  = 0$, which result from the mentioned  
invariant action $\int\limits_{surface} F$, are together with the  
two dimensionality condition: $d F (M_{2D}) = 0$ equivalent to the  
equations of motion in the four dimensional case: $d F^{\dagger}  
(M_{4D}) = 0$, $d F (M_{4D}) = 0$. So that in two dimensional case  
the relevant $F$ is given by $F \in Harm^2 (M_{2D})$ and in the four  
dimensional case the relevant $F$ is given by $F \in Harm^2  
(M_{4D})$. Nevertheless, as it is discussed above, the construction  
of the underlying four manifold is so that only the spatial two  
dimensional submanifold seems to be relevant for definition of the  
physical electromagnetic two-form. Thus in two dimensional case  
there is an isomorphism $H^0 (M_{2D}) \cong H^2 (M_{2D})$ which  
replaces the four dimensional isomorphism $H^2 (M_{4D}) \times H^2  
(M_{4D}) \rightarrow H^4 (M_{4D}) \cong H^0 (M_{4D})$.
Furthermore recall that the second cohomology of a four manifold  
$H^2 (M_{4D})$ is "destroyed" by removing the $M_{2D} \in H_2  
(M_{4D})$ surfaces from the four manifold $M_{4D}$ \cite{bott}.  
Therefore the Maxwell action $H^2 _{(em)} (M_{4D}) \times H^2  
_{(em)} (M_{4D}) \rightarrow H^4 _{(em)} (M_{4D}) \cong H^0_{(em)}  
(M_{4D})$ depends entirely on the two dimensional submanifold $  
M_{2D} \in H_2 (M_{4D})$. Accordingly also the constructing two-fom  
$F \in H^2 _{(em)} (M_{4D})$ depends only on the surface $ M_{2D}  
\in H_2 (M_{4D})$ and "effectively" it should be defined on such a  
surface.

Note also that the usual four dimensional coupling term $\int  
A_{\mu} j^{\mu} dt$ with $j^{\mu} = n e \dot{x}^{\mu}$ where $n$ is  
the electronic density, is equal to $Q \int A_{\mu} dx^{\mu}; \ Q =  
n e$ which reduces in the two dimensional case to $Q \int A_m dx^m$.  

Therefore the whole four dimensional action $\int\limits_{M_{4D}} F  
\wedge * F + \int A_{\mu} \cdot j^{\mu} dt$ reduces in the two  
dimensional single electron case to $e \oint A_m dx^m =  
\int\limits_{surface} F$, since in this case also $\int\limits_{4D}  
F \wedge * F$ reduces to $\int\limits_{M_{2D}} F$. Recall that in  
two dimensions $* \Omega^2 = \Omega^0$, then in view of $\Omega^0  
\cdot \Omega^2 \sim \Omega^2$ one has in this case $F (M_{2D})  
\wedge * F (M_{2D}) \sim F (M_{2D})$.

\bigskip
Therefore, in order to adopt all these facts in the theory, we  
conjecture a two dimensional invariant of two-form $F$, i. e.  
$\int\limits_{surface} F$, for the electromagnetic action which  
avoids problems with extra conditions for renormalization and with  
constraints of a four dimensional QED \cite{const}.

\bigskip
The two dimensional electromagnetic action of interest is given by  
the classical flux function

$S_{(cl)} = \Phi_{(cl)} = e \oint\limits_{(contour)} A_m dx^m = e  
\int\limits_{(surface)}  F_{mn} dx^m \wedge dx^n$, where $e$, $A_m$  
and $F_{mn}$ are, respectively, the electric charge of the electron,  
the electromagnetic potential and the magnetic field strength  
interacting with the  electron and $m, n = 1, 2$.
Here the domain of electromagnetic potential and magnetic field  
strength is a non-simply connected region containing of two regions  
which is similar to the case of Aharonov-Bohm effect: One is the  
flux {\it surface} where a constant magnetic field $B_{(surface)} =  
B_{(constant)} := \epsilon_{mn} F^{mn} _{(surface)}$ is present and
$A^m _{(surface)} = B_{(constant)} x_n \epsilon^{mn}$. The second  
is the {\it contour} region which surrounds this flux surface where  
the magnetic field is absent $B_{(contour)} =0$ and $A^m  
_{(contour)} = \partial^m \Phi, i. e.
d A_{(contour)} = B_{(contour)} = 0$. Thus, the integral $e  
\oint\limits_{(contour)}  A_m dx^m$ is defined on the contour  
region, whereas the equivalent integral $e \int\limits_{(surface)}  
F_{mn} dx^m \wedge dx^n$ is defined on the surface region.

\bigskip
We will prove that the canonical quantization of $S_{(cl)}$ is  
given by the  commutator postulate:

$e [ \hat{A}_m  ,  \hat{x}_m ] = -i\hbar$ which is related with a  
new uncertainty relation $e \Delta A_m \cdot \Delta x_m \geq \hbar$  
for photon \cite{qml}. Hereby functions $x_m$ are the position  
coordinates of an electron interacting with the magnetic field  
$F_{mn}$ of photon $A_m$. In other words in this approach photon is  
considered, in accord with the equivalence between quantum fields  
and quantum particles in quanrum field theory, as a quantum particle  
with usual uncertainty properties in measurments (see below).

With respect to the commutator postulate note that, in view of $A_m  
^{(surface)} = B_{(constant)} \cdot x^n \epsilon_{mn}$ and $A_m  
^{(contour)} := \partial_m \Phi$, in both relevant regions $A_m$ is  
not a function of $x_m$ and so there is no a periori reason for the  
commutativity of $\hat{A}_m$ and $\hat{x}_m$ operators.
Further recall that although the $A_m$ potential is non-observable  
in view of its gauge dependence, however the quantized integral $e  
\oint\limits_{(contour)}  A_m dx^m$ is in view of Aharonov-Bohm  
effect or flux quantization an observable phase. Thus the difference  
of two gauge potential $\Delta A_m \sim (\delta A_m = \tilde{A}_m -  
A_m)$ is, in view of gauge transformations: $A_m ^{\prime} = A_m +  
\partial_m \Lambda$, $\tilde{A}_m ^{\prime} = \tilde{A}_m +
\partial_m \Lambda$, a gauge invariant quantity.

Not that the quantization postulates $S_Q = e  
\oint\limits_{(contour)}  A_m dx^m = N h$, $e [ \hat{A}_m  ,   
\hat{x}_m ] = -i\hbar$ and

$e \Delta A_m \cdot \Delta x_m \geq \hbar$ can be compared with the  
canonical quantization postulates $\oint P_m dq^m = N h$, $[  
\hat{P}_m  ,  \hat{q}_m ] = -i\hbar$ and $\Delta P_m \cdot \Delta  
q_m \geq \hbar$.
Thus the quantization $S_Q = e \int\limits_{(surface)}  F_{mn} dx^m  
\wedge dx^n = N h$ is just the integrality condition for the first  
Chern class $ch_1 \sim F$ and in this sense it is a well defined  
geometric quantization \cite{wood}.

\bigskip

We will show that, indeed for the canonical conjugate variables of  
phase space of the two dimensional electromagnetic system which is  
represented by $S_{(cl)}$, the commutator of related operators is  
non-trivial. The key point is the correct choise of phase space, i.  
e. the choise of true canonical conjugate variables for the two  
dimensional electromagnetic system under consideration.

\bigskip
The point of departure is the two dimensional electromagnetic  
action functional:

\begin{equation}
S_{(cl)} = \Phi_{(cl)} = e \int\limits_{(surface)}  F_{mn} dx^m  
\wedge dx^n = e \int \limits_{(surface)}  dA_n \wedge dx^n = e  
\oint\limits_{(contour)}  A_n dx^n \ ,
\end{equation}
\label{one}

with $dA_n := \partial_m A_n dx^m \epsilon_{mn}$.

The action is defined on the electromagnetic $U(1)$ bundle over the  
two dimensional manifold which consists of a two dimansional  
non-interacting electronic system in magnetic field in the "single  
electron picture".

First we show that $S_{(cl)} = e \int\limits_{(surface)}  F_{mn}  
dx^m \wedge dx^n$ is a well defined action functional from which one  
can derive the equations of motion for $A_m$, so that it can be  
quantized canonically in order to describe the quantized dynamics of  
the two dimensional electromagnetic system.

In view of the fact that $A_m$ depends on $x^n$ by $A^m  
_{(surface)} = B_{(constant)} x_n \epsilon^{mn}$, the variation of  
action $\delta S_{(cl)}$ needs to be considered only with respect to  
the variation of $\delta x^n$, since the variation $\delta A_m$ is  
proportional to $\delta x_n$. Hence, one has to consider $dx^l =  
\displaystyle{\frac{\partial x^l}{\partial x^m}} dx^m$.
The Euler-Lagrange equations $\displaystyle{\frac{\partial  
L}{\partial x^m}} = \partial _n {\frac{\partial L}{\partial  
\partial_n x^m}}$ which result from the variation of this action  
with respect to the variation $\delta x^n$ are:

\begin{equation}
\partial_n \partial_n A_m - \partial_n \partial_m A_n = 0
\end{equation}

These are the usual equations of motion for $A_m$ potential in  
vacuum. Nevertheless, in view of the fact that
$A^{(surface)} _m = B_{(constant)} x^n \epsilon_{mn}$, the second  
term is identically zero and one is left with the Laplace equation  
in two dimensions $\partial^n \partial_n A_m = 0$ \cite{andere}.  
Note that also Maxwell equations in vacuum together with Lorentz  
condition $d^{\dagger} A = 0$ result in the four dimensional Laplace  
equation $d d^{\dagger} A + d^{\dagger} d A = 0$.

Therefore, the action $S_{(cl)} = e \int\limits_{(surface)}  F_{mn}  
dx^m \wedge dx^n = e \oint\limits_{(contour)}  A_m dx^m$ is a well  
defined action functional for our two dimensional system, which can  
be canonically quantized in order to describe the quantum behaviour  
of photon.

\bigskip
To quantize the phase space of a classical system which is  
represented by an action functional $S_{(cl)}$, one should determine  
first the canonical conjugate variables of phase space and then one  
should postulate the quantum commutator for operators which are  
related to these variables. Now to determine the phase space space  
variables of the system represented by the action functional  
$S_{(cl)}$ one can use the Legendre transformation formula $P_m :=  
\displaystyle{\frac{\partial L}{\partial \dot{q}}}$ which is defined  
for the phase space of canonical action: $\oint\limits_{phase \;  
space}  P_m \dot{q}^m dt$. Thus the phase space of our system which  
is represented by the action $S_{(cl)} = e \oint\limits_{(contour)}   
A_m dx^m = e \oint\limits_{(contour)}  A_m \dot{x}^m dt$ has the  
canonical conjugate variables ${\{ A_m, x^m}\}$ and it can be  
quantized directly in comparision with the phase space of canonical  
action as mentioned above. Nevertheless to be precize we perform the  
quantization of this system in accord with the general formalism of  
geometric quantization \cite{wood}:

Then, the globally Hamiltonian vector fields of our system with the  
symplectic two-form:

$\omega =  dA_n \wedge dx^n =  F_{mn} dx^m \wedge dx^n$ are given  
by the following differential operators \cite {wood}, \cite{erk}:

\begin{equation}
X_{A_m} = \displaystyle{\frac{\partial}{\partial x^m}} \;\;\; ,  
\;\;\; X_{x^m} = - \displaystyle{\frac{\partial}{\partial A_m}}
\end{equation}
\label{five}

Moreover, the quantum differential operators on the quantized phase  
space of this system should be proportional to these vector fields  
by a complex factor, i. e. usually by $(-i \hbar )$, and so they  
should be given by:

\begin{equation}
\hat{A}_m = -i \hbar \displaystyle{\frac{\partial}{\partial x^m}}  
\; \;\; , \; \; \; \hat{x}_m = i \hbar  
\displaystyle{\frac{\partial}{\partial A_m}}
\end{equation}
\label{six}

On the other hand, the real quantized phase space of a quantum  
system should be polarized in the sense that the $\Psi$ wave  
function of system should be a function of only half of the  
variables of the original phase space \cite{wood}. This means that  
it is either in the $\Psi ( A_m , t)$- or in the $\Psi (x^m , t)$  
representation.
Then the quantum operators are given, respectively, by the set ${\{  
\hat{A}_m = A_m \ , \ \hat{x}_m = -i \hbar X_{x^i} =
i \hbar \displaystyle{\frac{\partial}{\partial A_m}}}\}$ or by the  
set ${\{ \hat{A}_m = -i \hbar X_{A_m} = -i  
\hbar\displaystyle{\frac{\partial}{\partial x^m}} \ , \ \hat{x}^m =  
x^m }\}$.
Thus in both representations the commutator between the quatum  
operators is given by:

\begin{equation}
e [ \hat{A}_m \ , \ \hat{x}_n ] \Psi = -i \hbar \delta_{mn} \Psi \; \;,
\end{equation}
\label{six}

which is gauge invariant \cite{Nn}. Equivalently in accord with  
quantum mechanics there is a true uncertainty relation:

\begin{equation}
e \Delta A_m \cdot \Delta x_m \geq \hbar
\end{equation}

Here $\Delta x_m$ is the position uncertainty of the electron  
observed by the light which is proportional to the wave length of  
light \cite{heitler1}.

This approach considers the photon as a {\it quantum particle} with  
its typical uncertainties; Since in the same way that a measurment  
of momentum or position of an electron needs its interaction with a  
photon, the measurment of electric field strength of a photon needs  
its interaction with an electron or with a charged test body ( see  
also Ref. \cite{heitler1}). Thus for a time dependent  
electromagnetic potential, e. g. $A_m = E_m \cdot t$, there is also  
an uncertainty relation for the electric field strength which is  
given by $e \Delta E_m \cdot \Delta t \cdot \Delta x_m \geq \hbar$.
Note also that a similar {\it quantum} relation exists also in the  
four dimensional QED, although it is introduced phenomenologically  
in addition to the usual canonical unceertainty relation of QED  
\cite{heitler1}. Rather this additonal quantum relation, i. e. $Q  
\Delta E_x \cdot T \cdot \Delta x \geq \hbar; \; Q = N^{\prime} e \;  
, \, N^{\prime} \in {\mathbf Z}$, is essential for the consistency  
of the usual uncertainty relations of QED \cite{heitler2}.  
Considering $A_x = E_x \cdot T$, it is obvious that the additional  
quantum relation in QED is the same as the canonically obtained  
uncertainty relation in the two dimensional model for $N^{\prime}$  
non-interacting electrons which has the action $N^{\prime} e  
\oint\limits_{contour} A_m dx^m$.

Moreover considering the QED uncertainty relation $\Delta G_x \cdot  
\Delta x \geq \hbar$ where $G_i = \int d^3 x \epsilon_{ijk} E_j B_k  
; i, j, k = 1, 2, 3$ is the momentum of light which observes the  
position $x$ of electron \cite{heitler1}; a comparison with  
additional QED quantum relation $Q \Delta E_x \cdot T \cdot \Delta x  
\geq \hbar$ shows that: $\Delta G_x = Q \Delta E_x \cdot T = Q  
\Delta A_x$. Thus, in two dimensions, the momentum of four  
dimensional QED $G_m$ can be identified with the momentum of the two  
dimensional model of quantum electrodynamics $ Q A_m$. This shows  
the compatibility of four and two dimensional quantum models with  
respect to the momentum structure. To underline this property note  
that using Gauss's law for closed surfaces and $A_i = \epsilon_{ijk}  
x_j B_k$ for solutions of Maxwell equations, i. e. for constant  
$B_k$, the momentum in four dimensional case $G_i = \int d^3 x  
\epsilon_{ijk} E_j B_k$, is given by $G_i = Q A_i$. Hence the  
canonical identification of momentum of {\it two dimensional} photon  
with $A_m$, as it is performed above, is in agreement with the  
momentum concept in four dimensional QED. Although the four  
dimensional theory do not present any canonical conjugate position  
variable for this momentum. At any case the momentum of photon $A_m$  
is correlated with the momentum of electron $P_m$. Thus the {\it  
flat} connection $A^{(contour)} _m$ on the contour region is  
defined, in accord with
$F^{(contour)} _{mn} = 0$, by $(- i \hbar \partial_m - e A _m) \Psi  
= 0$ where $\Psi$ is the wave function of electron. Hence, in view  
of $e \Delta A_x = \Delta P_x$ in accord with $P_m \Psi = e A_m  
\Psi$, the momentums as well as uncertainty relations for  
interacting electron and photon becomes correlated as expected, e.  
g. in Ref. \cite{heitler1}.

Recall however that whereas in the two dimensional approach the  
uncertainty relations $e \Delta A_m \cdot \Delta x_m \geq \hbar$  
result directly from the canonical quantization of the action, in  
the four dimensional QED one needs various assumptions to introduce  
the " inaccuracy"  relation:
$Q \Delta E_x \cdot \Delta x \cdot T \geq \hbar$ which is not  
deducable from the quantum structure of the four dimensional QED  
\cite{heitler1}.
This seems to be an advantage of the two dimensional approach to  
quantum electrodynamics.

Moreover, in accord with $A^{(surface)} _m = B_{(constant)} x^n  
\epsilon_{mn}$ or with $\Delta A_m = B \cdot \Delta x^n  
|\epsilon_{mn}|$ there should be also an equivalent uncertainty  
relation which is given by: $e B \Delta x_m . \Delta x_n \geq \hbar  
|\epsilon_{mn}|$ , i. e. for $m \neq n$. This uncertainty relation  
is related to the quantum commutator postulate
$e B [ \hat{x}_m \ , \hat{x}_n ] = -i \hbar \epsilon_{mn}$ which is  
equivalent to the quantization postulate: $S_{(cl)} =  
eB_{(constant)} \epsilon_{mn} \int\limits_{(surface)} dx^m \wedge  
dx^n = Nh$.
Recall that the commutator $e B [ \hat{x}_m \ , \hat{x}_n ] = -i  
\hbar \epsilon_{mn}$ is known, phenomenologically, as the commutator  
of relative electron coordinates operators in the cyclotron motion  
\cite{aoki} and it seems to be related with the so called "Peierls  
substitution" for electrons in strong magnetic fields \cite{pei}.
Furthermore in accord with the uncertainty relations the quantized  
electromagnetic gauge potential possess a maximal uncertainty of  
$\Delta A_m = (\Delta A_m)_{(maximum)} =  
\displaystyle{\frac{\hbar}{e l_B}}$ for the case $\Delta x_m =  
(\Delta x_m)_{(minimum)} = l_B$. Using $\Delta A_m = e B \cdot l_B $  
one obtains from uncertainty relation the independent  
phenomenological definition of magnetic length: $ l^2 _B =  
\displaystyle{\frac{\hbar}{e B}}$, which proves the consistency of  
this approach.
Therefore this two dimensional quantum theory which is represented  
by the equivalent canonical quantum postulates: i. e. by $e  
\oint\limits_{(contour)}  A_m dx^m =  e \int\limits_{(surface)}   
F_{mn} dx^m \wedge dx^n = \Phi_{(Q)}  = N h , N \in {\mathbf Z}$  ;
$e [ \hat{A}_m \ , \ \hat{x}_n ] = -i \hbar \delta_{mn}$ or $e B [  
\hat{x}_m \ , \hat{x}_n ] = -i \hbar \epsilon_{mn}$ can be  
considered as an appropriate theory to describe two dimensional  
quantum effects in presence of magnetic fields.

\bigskip

In conclusion let us remark that in accord with this canonical  
quantized model there should be a quantum of length equal to the  
magnetic length $l_B$ in two dimensional quantum electrodynamical  
systems, which was introduced phenomenologically in the magnetic  
quantization.
Accordingly in such quantum systems all relevant lengths and areas  
are quantized in units of $l_B$ and $l_B ^2$, respectively, since  
$(\Delta x_m)_{(minimum)} = l_B$ and $(\Delta x_m \cdot \Delta  
x_n)_{(minimum)} = l_B ^2$. Furthermore the introduced quantum  
commutator postulates $e [ \hat{A}_m \ , \ \hat{x}_n ] = -i \hbar  
\delta_{mn}$ or
$e B [ \hat{x}_m \ , \hat{x}_n ] = -i \hbar \epsilon_{mn}$ are  
equivalent, in accord with $A^{(surface)} _m = B_{(constant)} x^n  
\epsilon_{mn}$, to the commutator $e [ \hat{A}_m \, , \hat{A}_n ] =  
-i \hbar B$ which is equivalent to the quantum commutator postulate  
for the $2+1$ dimensional Chern-Simons theory in the $A_t = 0$ gauge  
\cite{witten}.
This circumstance relates our gauge free model with the {\it  
gauged} Chern-Simons models of topological field theories which are  
used also to describe QHE \cite{qhe}. Moreover the present model  
seems to be related, by quantum postulate $e B [ \hat{x}_m \ ,  
\hat{x}_n ] = -i \hbar \epsilon_{mn}$, to the Chern-Simons quantum  
mechanics \cite{jakiw}.


\bigskip
\begin{thebibliography}{100}



\bibitem{nak}
For general aspects of topology in physics see: M. Nakahara:  
"Geometry, Topology And Physics" (Adam Hilger, 1990).


\bibitem{dirac}
P. A. M. Dirac, V. A. Fock and Boris Podolosky (1932): in in  
"Selected Papers on Quantum Electrodynamics", edited by: J.  
Schwinger, (Dover Publications, Inc. New York, 1958).

\bibitem{bound}
As long as there is no {\it quantum prescription} to define or to  
measure the boundary of a manifold under quantum conditions, a  
general manifold should be considered as boundaryless.

\bibitem{weis}
V. Weisskopf, Danske Vid., (1936): in "Selected Papers on Quantum  
Electrodynamics", edited by: J. Schwinger, (Dover Publications, Inc.  
New York, 1958).


\bibitem{land}
L. D. Landau, E. M. Lifschitz, III Vol. We mean here the  
Landau-gauge $A_m := \epsilon_{mn} B \cdot x^n$;

$\epsilon_{mn} = - \epsilon_{nm} = -1$ where $B$ is a {\it  
constant} magnetic field: $B = \epsilon_{mn} F^{mn}, dB = 0$.

\bibitem{bohr}
N. Bohr and L. Rosenfeld, Phys. Rev., 78, 794, (1950).

\bibitem{bott}
R. Bott, Canad. Math. Bull. Vol. 28(2), 1985.

\bibitem{const}
Obviously the main constraint in a general four dimensional field  
theory is due to the fact that the introduced time component of the  
"field" is no dynamical variable, therefore such a constraint can be  
avoided in the absence of such a time component.


\bibitem{qml}

We denote functions on the phase space by $ x \;, \; A$, etc. and  
related quantum operators on the quantized phase space by $\hat{x}  
\;, \; \hat{A}$, etc. and set the velocity of light $C = 1$.  
Although the model is performed for a single electron, however it  
can be applied also to non-interacting two dimensional electronic  
system where the "single electron picture" is available.


\bibitem{wood}
N. Woodhouse,"Geometric Quantization",  (Clarendon Press, 1980,  
1990) Oxford University.





\bibitem{andere}
The resulting equations of motion from $S_{(cl)} =  
\oint\limits_{(contour)} e A_m dx^m$, i. e. $\partial_m A^m = 0 \sim  
d^{\dagger} A = 0$ together with the contour condition $d A = 0$,  
are equivalent to Laplace equation $\partial_n \partial^n A_m = (d  
d^{\dagger} + d^{\dagger} d) A = 0$.




\bibitem{erk}
According to Ref. \cite{wood} the classical Hamiltonian vector  
fields related eith the canonical conjugate variables ${\{ A_m , x^m  
}\}$ are given in general by:

$X_{A_m} = \displaystyle{{\frac{\partial A_m}{\partial  
A_n}}{\frac{\partial}{\partial x^n}} - {\frac{\partial A_m}{\partial  
x^n}}{\frac{\partial}{\partial A_n}}}$ \qquad, \qquad
$X_{x^m} = \displaystyle{{\frac{\partial x^m}{\partial  
A_n}}{\frac{\partial}{\partial x^n}} - {\frac{\partial x^m}{\partial  
x^n}}{\frac{\partial}{\partial A_n}}}$

Furthermore the inner product of any globally Hamiltonian  
vectorfield $X_f$ with symplectic two-form of the system, i. e.  
$\omega$, should result in: $< X_f , \omega > = - df$.



\bibitem{Nn}
In accord with Refs. \cite {wood} and \cite{erk} the vector field  
of $d \lambda$ have vanishing inner product with the symplectic  
two-form of the system, i. e. $< X_{d \Lambda} , \omega > = - d^2  
\Lambda \equiv 0$. The corresponding operator is then a constant  
operator proportional to the identity operator which commutes with  
all others.

\bibitem{heitler1}
W. Heitler: The Quantum Theory of Radiation, Third Edition, (Dover  
Publications, Inc. New York 1984): (II-9 and II-7).


\bibitem{heitler2}
Recall that the "charged test body" of Ref. \cite{heitler1} obeys  
the same uncertainty relations as the usual electron. Moreover the  
discussed two dimensional quantum electrodynamics is not changed, if  
the constant $e$ is replaced by $Q$.

\bibitem{aoki}
H. Aoki: Rep. Prog. Phys. 50, (1987), 655.


\bibitem{pei}
R. Peierls, Z. Phys. 80, 763 (1933).


\bibitem{witten}
E. Witten, Cumm. Math. Phys. 121 (1989) 351-399.


\bibitem{qhe}
G. W. Semenoff, Phys. Rev. Lett., 61, 517 (1988);
S. C. Zhang, T. H. Hanssson and S. Kivelson, Phys. Rev. Lett. 62,  
82 (1989).



\bibitem{jakiw}
R. Jackiw, G. Dunne, Nucl. Phys. B (proc. suppl.) 33c, 114-118 (1993).

\end{thebibliography}
\end{document}